\date{\today}

        \documentclass[%
        reprint,
        superscriptaddress,
        amsmath,amssymb,
        pra,
        ]{revtex4-2}
        \usepackage{booktabs}
        \usepackage{graphicx}
        \usepackage[T1]{fontenc}
        \usepackage{graphicx}
        \usepackage{dcolumn}
        \usepackage{braket}
        \usepackage{bm}
        \usepackage[utf8]{inputenc}
        \usepackage{stmaryrd}
        \allowdisplaybreaks
        
\begin{document}
        	
        	\preprint{APS/123-QED}
        	
        	\title{Long-distance quantum secure direct communication  with Gottesman-Kitaev-Preskill qubits}
        	
        	\author{Meng-Dong Zhu}
        	\affiliation{College of Electronic and Optical Engineering, \& College of Flexible Electronics (Future Technology), Nanjing University of Posts and Telecommunications, Nanjing, 210023, China}
        	\author{Jia-Wei Ying}
        	\affiliation{College of Electronic and Optical Engineering, \& College of Flexible Electronics (Future Technology), Nanjing University of Posts and Telecommunications, Nanjing, 210023, China}
        	\author{Qi Zhang}
        	\affiliation{College of Science, Nanjing University of Posts and Telecommunications, Nanjing 210023, China}
        	\author{Shi-Pu Gu}
        	\affiliation{College of Electronic and Optical Engineering, \& College of Flexible Electronics (Future Technology), Nanjing University of Posts and Telecommunications, Nanjing, 210023, China}
        	\author{Xing-Fu Wang}
        	\affiliation{College of Science, Nanjing University of Posts and Telecommunications, Nanjing 210023, China}
        	\author{Lan Zhou}
        	\email{zhoul727@hznu.edu.cn}
        	\affiliation{School of Physics, Hangzhou Normal University, Hangzhou,  311121, China}
        	\author{Yu-Bo Sheng}
        	\email{shengyb@hznu.edu.cn}
        	\affiliation{School of Physics, Hangzhou Normal University, Hangzhou,  311121, China}

        	\date{\today}

\begin{abstract}
The Gottesman-Kitaev-Preskill (GKP) code possesses a lattice periodic structure that enables the effective correction of small displacement errors induced by noise in phase space, which is indispensable for quantum computation. The GKP code was also used  to construct a long-distance entanglement channel in quantum repeaters and distribute random keys in quantum key distribution (QKD). Using GKP encoding, we propose a quantum secure direct communication (QSDC) protocol that not only transmits the secret message directly, but also helps protect the safety of communication personnel. 
The exponential attenuation loss of photons can be overcome by the third-generation quantum repeater. 
Numerical simulation results demonstrate that secure transmission distance exceeding 1000 km is achievable with a 23 dB squeezing level. 
Furthermore, we establish a cost-based optimization framework to characterize the trade-off among secure communication distance, secrecy capacity, and repeater resources, enabling optimized repeater deployment and providing a systematic design strategy for scalable quantum repeater architectures.
The results  demonstrate the potential of combining GKP error correction for long-distance QSDC and quantum networks.
\end{abstract}

\maketitle


\section{\label{sec.1}Introduction}
Quantum error correction (QEC) is essential for protecting quantum information from noise, providing a foundation for fault-tolerant quantum computation \cite{steaneQEC,calderbankShor,knillLaflamme,surfacecode,newadd1,newadd2}, continuous displacement sensing \cite{sensing} and long-distance quantum communication \cite{qecquantumcommunication}. Bosonic codes \cite{bosonic1,bosonic2,bosonic3,bosonic4,bosonic5} have emerged as a promising class of QEC codes because they exploit the infinite-dimensional Hilbert space of an oscillator to encode quantum information redundantly within a single physical mode.
 Gottesman-Kitaev-Preskill (GKP) code \cite{GKP}, proposed in 2001, can enable effective correction of small displacement errors in the quadrature components \cite{GKPerrorAnalysis}. Its compatibility with Gaussian operations allows efficient Clifford gates, syndrome extraction, and error correction \cite{allGaussianGKP}, while the analog syndrome information can further improve decoding and facilitate concatenation with discrete-variable codes \cite{analog-threshold,surfaceGKP,lowOverheadSurfaceGKP}. These features make GKP encoding a promising route toward fault-tolerant quantum information processing \cite{cvMBQCArchitecture,bosonic5,gaussian1,gaussian2}.

Currently, the generation and manipulation of GKP qubits have been achieved in superconducting systems \cite{superconduct1,superconduct2}, trapped-ion systems \cite{ion,iongateachieve} and integrated photonics \cite{integratedGKP2025}. In optical systems, the preparation of non-Gaussian states remains challenging due to the requirement for strong nonlinear interactions. Several novel schemes for the generation of GKP states have been proposed, including breeding methods utilizing non-Gaussian resource states \cite{breeding1,breeding2}, Gaussian boson sampling based on photon-number-resolving detectors introducing nonlinearity \cite{bosonsampling1,bosonsampling2}, and measurement-free approaches \cite{measurementfree}. 
Furthermore, a recent all-optical proposal utilizes Gaussian inputs, optical parametric amplification, and heralded photon detection to generate squeezed-cat states that can be bred into approximate GKP states \cite{2026opt}.
In addition to physical generation schemes, neural-network-based optimization has recently been proposed to design approximate GKP codes with reduced state complexity while preserving error-correction performance \cite{neuralnetwork}.  In 2024, the first experimental generation of a GKP-like quantum optical state in a propagating wave system was demonstrated \cite{opticalGKP} using cat states and homodyne measurements. Notably, the GKP state was exploited in long-distance quantum key distribution (QKD) \cite{thirdrepeater2,concatenated1,concatenated2,thirdrepeater2,highdimensionGKP1}, combined with the third-generation quantum repeater protocols \cite{highdimensonGKP2,quantumrepeater2,thirdrepeater1}.

QKD combined with a one-time pad provides a promising method to realize secure communication \cite{qkdgisin}. Beyond QKD, quantum secure direct communication (QSDC) enables the secret message to be transmitted directly without prior shared keys, while also allowing the detection of an eavesdropper (Eve) before information is leaked \cite{entanglement_QSDC,DL04,qsdcprl,qsdcreview,qkdbeyond,QSDC_NSR2025}.  
In recent years, QSDC has progressed rapidly from theoretical protocol design to experimental demonstrations. 
Advanced schemes such as measurement-device-independent QSDC \cite{MDIQSDC1,MDIQSDC2,SPMDIQSDC2024}, receiver-device-independent QSDC \cite{RDIQSDC2025} and device-independent (DI) QSDC \cite{DIQSDC} have strengthened its security framework. Notably, recent work on GKP-state Bell nonlocality \cite{GKPMABK} reveals the restrictions of Bell-pair states and the strong Bell violations achievable with multipartite GKP-encoded states under homodyne detection, offering insights for DI-based GKP communication.
Meanwhile, the one-step QSDC protocols \cite{onestepQSDC1,onestepQSDC2,onestepQSDC3} can mitigate message loss caused by photon transmission loss and high-dimensional QSDC can increase the information capacity per quantum carrier by exploiting larger Hilbert spaces \cite{highdimQSDC2024,HDOPIQSDC2025}.
Passive source-preparation schemes further improve practical security against source-side imperfections \cite{passiveDecoyQSDC2024,passiveStateQSDC2025}.
Meanwhile, both discrete-variable and continuous-variable QSDC have been experimentally demonstrated \cite{experiment1,qsdcprl,experiment4,secureRelayQSDC2023,practicalQSDC2019,QSDCnetwork2021,OAM_CVQDC2025,coherentQSC2026,cvQSDC_PRApp2021,fiberQSDC2022,freeSpaceQSDC2020}, with transmission distances extended from 100 km in fiber in 2022 \cite{experiment3} to a recent four-user fully connected QSDC demonstration over 300 km of total fiber transmission \cite{2025_300km}.
Moreover, by detecting eavesdropping in real-time during communication, such protocols \cite{entanglement_QSDC,DL04,qsdcprl,qsdcreview}  provide a powerful approach to protecting the safety of communicators. Once eavesdropping is detected during the communication process, both parties not only stop communication, but also immediately relocate and hide the communication personnel.
It is extremely useful to avoid decapitation strikes on communicators caused by exposing communication positions in modern warfare.
In such quantum communication protocols, the secret message is encoded in quantum states or entanglement \cite{entanglement_QSDC,DL04,qsdcprl}.  Unfortunately, in long-distance information transmission, the first-generation  quantum repeater \cite{quantumrepeater1} is  not suitable for QSDC protocols, as entanglement purification will erase the encoded information after purification. An alternative approach is to exploit QEC codes, such as repetition \cite{qsdcding} and GKP codes \cite{GKPQldpc}, respectively. Without quantum repeaters, the above protocols are limited to point-to-point communication, and they cannot achieve longer-distance QSDC or scalable networks.

In this protocol, combined with the third-generation quantum repeater \cite{quantumrepeater2,thirdrepeater1,thirdrepeater2}, we propose the long-distance QSDC protocol using GKP qubits. We show that the third-generation quantum repeater is extremely suitable for direct transmission of secure information. At each repeater node, optical amplification, continuous-variable SUM gates and ancillary GKP states are used to perform error correction, thereby suppressing the noise accumulated from channel loss and imperfect state preparation. By tracking the variance evolution across a repeater chain, we obtain the logical error probabilities and derive a lower bound on the secrecy message capacity. Numerical results show that, for a GKP squeezing level of about 23 dB and 2750 repeater nodes, secure communication can be extended beyond 1000 km.
Furthermore, we develop a cost-based optimization framework that establishes a quantitative relationship between communication distance, secrecy capacity, and repeater resources. Under the minimized cost function, high-quality information transmission with a secrecy message capacity of 0.70 bit/state can be achieved over a distance of 3.46 km using one repeater node.
These results highlight the potential of integrating GKP error correction to facilitate long-distance QSDC.

This paper is organized as follows. In Sec. \ref{sec.2}, we review the fundamental properties of GKP qubits, the photon-loss channel, and the error-correction procedure. In Sec. \ref{sec.3}, we describe the proposed QSDC protocol in detail, including the structure of the repeater nodes, the encoding and decoding process. Sec. \ref{sec.4} presents the security analysis and the model to estimate the secrecy message capacity. In Sec. \ref{sec.5}, we provide numerical simulation results. We discuss the potential realization of our protocol in Sec. \ref{sec.6}. Finally, Sec. \ref{sec.7} concludes the paper with a summary and outlook for future work.

\section{\label{sec.2}Background}
In this section, we review the GKP qubit, the photon-loss channel, amplification and the single-qubit quantum error correction.

\subsection{\label{2.1}The GKP qubit}
First, we consider a harmonic oscillator with annihilation and creation operators $\hat{a}$ and $\hat{a}^{\dagger}$ satisfying $[\hat{a},\hat{a}^{\dagger}]=1$. The position and momentum operators of a bosonic mode are defined as $\hat{q}=(\hat{a}+\hat{a}^{\dagger})/\sqrt{2}, \hat{p}=i(\hat{a}^{\dagger}-\hat{a})/\sqrt{2}$. Their eigenstates, denoted as $|q\rangle, |p\rangle$, are called quadrature states and are related through the Fourier transform $|q\rangle = \int \frac{dp}{\sqrt{2\pi}} e^{ipq} |p\rangle$. One of the ideal GKP qubit is to encode quantum information in translation-invariant grid states, which is stabilized by the two stabilizers \cite{GKP}
\begin{equation}
\hat{S}_q=exp(i2\sqrt{\pi}\hat{q}),\quad \hat{S}_p=exp(-i2\sqrt{\pi}\hat{p}).
\end{equation}
The stabilizers commute with each other and can be measured simultaneously.

The ideal GKP logical states in the Z basis, expressed in the $(q, p)$ phase space, are given by \cite{GKP}
\begin{align}
|0_{L}\rangle & = \sum_{n \in \mathbb{Z}}|q =2n\sqrt{\pi}\rangle, \\
|1_L\rangle & = \sum_{n \in \mathbb{Z}} |q = (2n + 1)\sqrt{\pi}\rangle.
\end{align}
Similarly, the states in the complementary basis are
\begin{align}
|+_{L}\rangle & = \sum_{n \in \mathbb{Z}}|p =2n\sqrt{\pi}\rangle, \\
|-_L\rangle & = \sum_{n \in \mathbb{Z}} |p = (2n + 1)\sqrt{\pi}\rangle.
\end{align}
All these basis states satisfy $\hat{q} \equiv 0$ and $ \hat{p}\equiv 0$ (modulo $\sqrt{\pi}$). Consequently, shift errors in the quadrature components $\hat{q}$ and $\hat{p}$ smaller than $\sqrt{\pi}/2$ can be corrected.
Pauli operators of the GKP qubit can be denoted by the square root of the two stabilizers, as follows
\begin{align}
    &\hat{Z}=exp(i\sqrt{\pi}\hat{q})\\
    &\hat{X}=exp(-i\sqrt{\pi}\hat{p}),
\end{align}
which can induce a logical phase-flip and bit-flip in the Z basis.

However, the ideal GKP qubit, which consists of infinitely squeezed states, requires infinite energy and is therefore unphysical and unrealistic. We assume that the states are composed of a series of Gaussian peaks of width $\triangle$ contained in a larger Gaussian envelope of width $1/\triangle$ \cite{thirdrepeater2}.
The states become the ideal GKP qubit as long as infinite squeezing $(\triangle \rightarrow 0)$. The value of $\triangle^2$ is set to $2\sigma^2$ such that the variance of each tooth in both quadrature components equals $\sigma^2$.

Like Refs. \cite{thirdrepeater2,concatenated1}, a simplified model that incorporates finite squeezing effects while neglecting the effect of the phase-space envelopes is adopted. The quadrature components are no longer modulated deterministically at points of odd or even multiples of
$\sqrt{\pi}$; instead, they result in probabilistic events following a Gaussian distribution centered at these multiples.
Accordingly, we assume that the probabilities of X and Z errors in the $\hat{p}$ and $\hat{q}$ quadrature components separately and estimate the probability of a logical GKP state error $P_{err}$ after error correction as follows
\begin{equation}\label{eq.8}
	\small
    P_{err}^X=P_{err}^Z=1-\int_{\frac{-\sqrt{\pi}}{2}}^{\frac{\sqrt{\pi}}{2}} dx\frac{1}{\sqrt{2\pi}\sigma_{final}}\exp\left(-\frac{x^{2}}{2\sigma^{2}_{final}}\right),
\end{equation}
which is an upper bound on the logical error probability due to the neglect of low-probability events where the displacement is modulated to points centered at other even multiples of $\sqrt{\pi}$, with a surrounding interval of $[-\sqrt{\pi}/2,\sqrt{\pi}/2)$.

\subsection{\label{2.2}The photon-loss channel and amplification}
Photon loss can be modeled by applying a beam splitter to the input state and a vacuum state.
After propagating through a loss channel of length $L$, corresponding to the repeater spacing, the variances in the $\hat{q}$ and $\hat{p}$ quadratures transform as follows \cite{thirdrepeater2}
\begin{equation}
    \sigma^2_{q,p}\rightarrow \eta \sigma^2_{q,p}+(1-\eta) \sigma^2_{vac},
\end{equation}
where $\eta$ includes the transmission efficiency of the channel $exp(-L/L_0)$ with the attenuation length $L_0$=22 km, the fiber coupling efficiency $\eta_0$ and the detection efficiency $\eta_h$ of the practical homodyne detectors. Here, $\sigma^2_{vac}$ is the variance of the vacuum state in the quadrature components $\hat{q}$ and $\hat{p}$, both of which are 1/2.

To partially compensate for the effects of the loss channel, one can apply an optical amplification operation. Under such amplification, the variances of the $\hat{q}$ and $\hat{p}$ quadratures transform as follows \cite{thirdrepeater2}
\begin{equation}
    \sigma^2_{q,p}\rightarrow \frac{1}\eta{} \sigma^2_{q,p}+ \frac{1-\eta}{\eta} \sigma^2_{vac}.
\end{equation}

Amplification can be applied before the one-way transmission, resulting in the following transformation of the variances of both the
$\hat{q}$ and $\hat{p}$ quadratures \cite{thirdrepeater2}
\begin{align}
        \sigma^2_{pre}= \sigma^2_{q,p}+ 1-\eta
\end{align}
It can be seen that the variance still increases slightly after transmission through this channel even with amplification.

\subsection{\label{2.3}The detail of the GKP qubit error correction}

The controlled‑X (CNOT) gate from control mode $D$ to target mode $A$ corresponds to the continuous‑variable SUM gate \cite{GKP,concatenated1}, which is required for the error correction.
\begin{equation}
CNOT_{D\rightarrow A}=SUM_{D\rightarrow A}=exp(-i\hat{q}_D\hat{p}_A)
\end{equation}
where $D$ denotes the data mode and $A$ refers to the ancilla mode. i.e.,
\begin{align}
&CNOT|s_1\rangle_{q_D}|s_2\rangle_{q_A} = |s_1\rangle_{q_D}|s_2 + s_1\rangle_{q_A}\\
&CNOT|t_1\rangle_{p_D}|t_2\rangle_{p_A} = |t_1 - t_2\rangle_{p_D}|t_2\rangle_{p_A}.
\end{align}

\begin{figure}[htbp]
\centering
\includegraphics[scale=0.5]{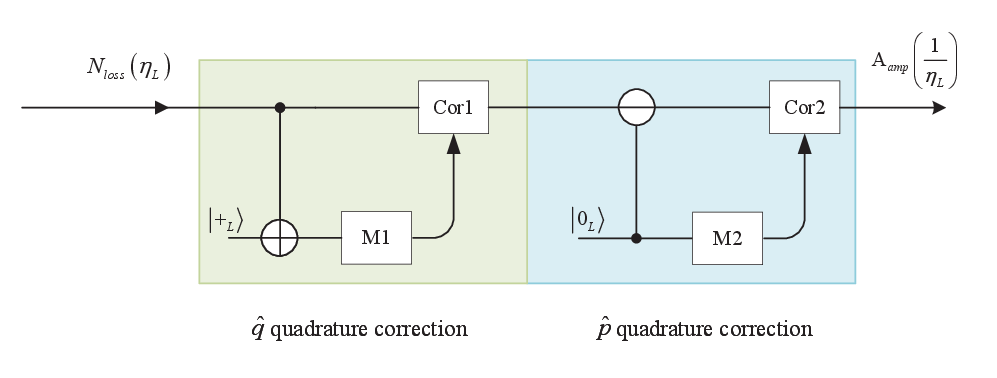}
\caption{The structure of the repeater with amplification and steane error-correction \cite{concatenated1}. The GKP state from the previous repeater node is sent through a bosonic pure-loss channel of distance $L$ with transmittance $exp(-L/L_0)$. Upon arrival at the repeater node, error correction is performed sequentially on the $\hat{q}$ and $\hat{p}$ quadratures, followed by optical amplification. The GKP state is then transmitted to the next node. }
\label{FIG.3}
\end{figure}

The error correction on GKP qubits is implemented as formulated in Ref. \cite{concatenated1}, which employs the SUM gates and ancilla GKP states. Specifically, the steane error-correction structure for the $\hat{q}$ and $\hat{p}$ components is shown in Fig. \ref{FIG.3}.

In the ideal scenario, the error affecting the data mode in the $\hat{q}$ and $\hat{p}$ quadrature components follows a Gaussian distribution with zero mean and variance $\sigma_{data}^2$.
Error correction for the GKP code is implemented by measuring the two stabilizers $\hat{S}_q$ and $\hat{S}_p$.
These measurements are realized by preparing an ancilla GKP state—either $|+_L\rangle$ for the $\hat{S}_q$ measurement or $|0_L\rangle$ for the $\hat{S}_p$ measurement—followed by applying the $\text{SUM}_{D\rightarrow A}$ or $\text{SUM}_{A\rightarrow D}^\dagger$ gate respectively. Subsequently, the $\hat{q}$ or $\hat{p}$ quadrature of the ancilla is measured via homodyne detection.
The measurement outcomes $(q_A, p_A)$ modulo $\sqrt{\pi}$ yield the estimated shifts $(q_{\text{err}}, p_{\text{err}})$, which correspond to the error to be corrected by applying a displacement to the data qubit. In order to simplify the analysis and to obtain a lower bound on the key rate of communication, as in Eq. \ref{eq.8}, we assume that any shift error lying outside the interval $[-\sqrt{\pi}/2,\sqrt{\pi}/2)$ results in a logical error.

Considering that the ancilla GKP states are also not prepared perfectly and have a variance of $\sigma_{GKP}^2$, implementing the displacement: $(q_D, p_D) \rightarrow (q_D - q_{err}, p_D -p_{err})$ introduces additional errors attributable to the ancilla states. Correspondingly, the variance of the data mode transforms from $(\sigma_{data,q}^2, \sigma_{data,p}^2)$ to $ (\sigma^2_{GKP}, \sigma^2_{GKP})$. We adopt the method proposed by Rozpedek et al. in Ref. \cite{concatenated1},
which introduces a scaling factor $c$ to improve the performance of the correction, and the optimal $c$ is $\frac{\sigma^2_{data}}{\sigma^2_{data} + \sigma^2_{GKP}}$, leading to the variance of error distribution from $\sigma_{data}^2$ to  $c_{opt}\sigma_{GKP}^2$. Since $c_{\text{opt}}$ lies in the interval $(0, 1)$, this enables a lower bit error rate and probability of error accumulation during multi-round error correction.

Owing to gate operations, channel loss, and imperfectly prepared ancillary states, the variance $\sigma^2_{fina}$ used in the logical error calculation via Eq. \ref{eq.8} is $2\sigma_{GKP}^2+1-\eta+c_{opt}\sigma_{GKP}^2$, and after error correction, the data qubit again follows a Gaussian distribution with variance $c_{opt}\sigma_{GKP}^2$.
Finally, we can compute the optimal $c$ and $\sigma^2_{final}$.

\begin{equation}
\fontsize{6pt}{7pt}\selectfont
    c_{opt} =
\frac{-\sigma_{GKP}^{2} - (1-\eta)
+ \sqrt{\left(\sigma_{GKP}^{2} + 1-\eta\right)
\left(5\sigma_{GKP}^{2} + 1-\eta\right)}}
{2\sigma_{GKP}^{2}} ,
\end{equation}

\begin{equation}
\fontsize{5.5pt}{7pt}\selectfont
\sigma_{final}^{2} = \frac{1}{2}\left(
3\sigma_{GKP}^{2} + 1-\eta
+ \sqrt{\left(\sigma_{GKP}^{2} + 1-\eta\right) \left(5\sigma_{GKP}^{2} + 1-\eta\right)}
\right).
\end{equation}

\section{\label{sec.3}the QSDC protocol using GKP qubits}
\begin{figure*}[htbp]
	\centering
	\includegraphics[scale=0.5]{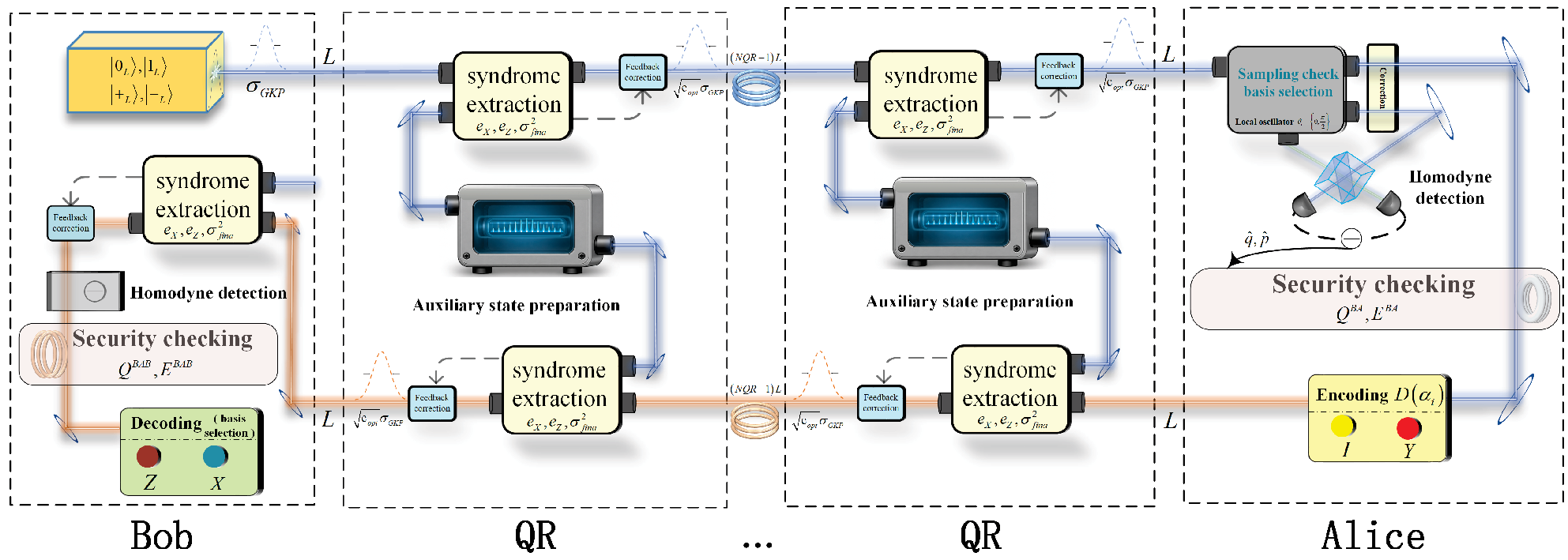}
	\caption{The structure of the QSDC protocol based on repeaters with GKP qubits.
	Bob prepares a sequence of finite-squeezing GKP states with the variance of $\sigma_{GKP}^{2}$ and transmits them to Alice through the forward quantum channel containing $NQR$ intermediate GKP repeater nodes. Alice performs error correction on the received modes. Upon passing the first security checking, she applies the encoding operation and sends the modes back to Bob through the backward quantum channel. Bob subsequently performs the final GKP
	error-correction and, after the second security checking, recovers the encoded message. For the translationally invariant repeater chain, each repeater resets the effective displacement-noise variance to $c_{opt}\sigma_{GKP}^{2}$, so that the logical bit- and phase-error probabilities are identical at all repeater nodes.
	$E^{BA}, E^{BAB}$ are the error rates of the Bob-Alice and the Bob-Alice-Bob channel; $Q^{BA}, Q^{BAB}$ are the overall signals of Alice and Bob; $e_X, e_Z$ are the probabilities of logical X and Z errors; $c_{opt}$, the optimal parameter; $\sigma_{final}^2$, the final variance during error correction.}
	\label{FIG.2}
\end{figure*}

The schematic of our protocol is shown in Fig. \ref{FIG.2}. This QSDC protocol incorporates the third-generation quantum repeater based on GKP qubits \cite{thirdrepeater1,thirdrepeater2}. It has the following steps.

Step 1:
Bob randomly selects the $Z$ or $X$ basis to prepare a sequence of GKP states $(|0_{L}\rangle, |1_{L}\rangle, |+_{L}\rangle, |-_{L}\rangle)$ and transmits them to Alice.

Step 2:
The states pass through quantum repeaters, each of which incorporates both $\hat{
p}$ and $\hat{q}$ quadrature error correction structures, corresponding to the use of auxiliary logical $|0_{L}\rangle$ and $|+_{L}\rangle$ states for error correction. The error is transferred from the data mode to the ancilla mode, and then a feedback displacement is applied to the corresponding quadrature of the data mode based on homodyne detection of the ancilla. After this feedback operation, an amplification operation is performed on the data mode to better compensate for channel loss.

Step 3:
Alice then receives the states and randomly selects some states as samples for detecting eavesdropping. She randomly performs error correction in $\hat{q}$ or $\hat{p}$ quadrature component and makes corresponding syndrome measurements (homodyne detection) on these samples.
She then announces the positions of the samples, along with the measurement basis (i.e., $\hat{q}$ or $\hat{p}$) and homodyne outcomes. Alice and Bob retain only the samples where their chosen bases coincide and get the detection bit error rate (DBER) through a classical authenticated channel.

\begin{figure*}[htbp]
	\centering
	\includegraphics[scale=0.85]{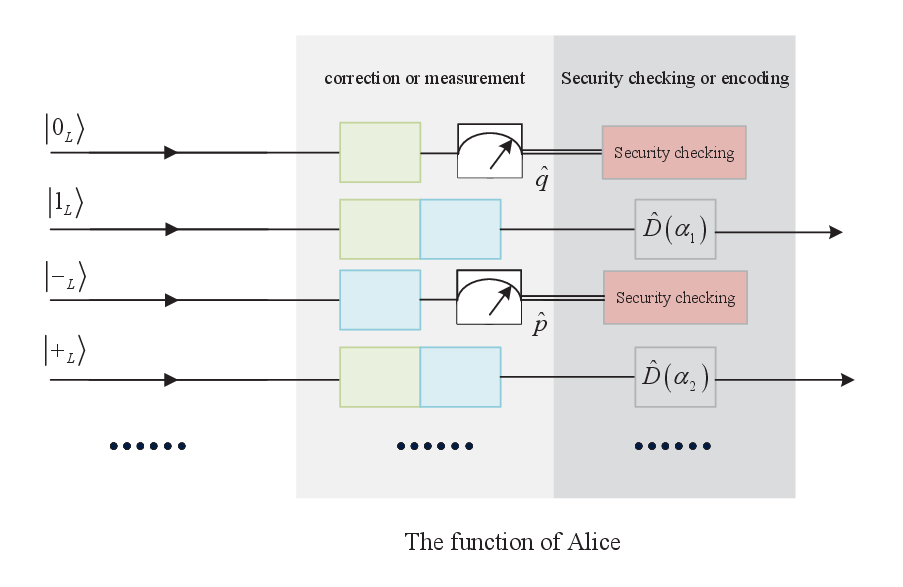}
	\caption{Schematic diagram of the error correction and encoding process performed at Alice’s side. Alice randomly selects some of states to detect eavesdropping. Specifically, for each selected state, error correction and measurement are performed on either the $\hat{q}$ or $\hat{p}$ quadrature. For the remaining states, both quadratures are simultaneously error-corrected, and encoding is carried out using displacement operators. $D(\alpha_i)$ is the displacement operation and $\alpha_i \in \{0, (\sqrt{\pi}+i\sqrt{\pi})/\sqrt{2}$\}, $(i=0, 1)$.
	}
	\label{FIG.4}
\end{figure*}

Step 4:
If the DBER is below a predetermined threshold, Alice applies the same error correction operation as implemented at the repeater node to the remaining states. As illustrated in Fig. \ref{FIG.4}, Alice performs displacement operation $\hat{D}(0)$ or $\hat{D}((\sqrt{\pi}+i\sqrt{\pi})/\sqrt{2})$ 
to encode the secret information bit 0 or 1, and returns the states to Bob through repeaters. Additionally, she randomly encodes some states for error-checking.

Step 5:
After receiving the sequence of states, Bob performs error correction on the quadrature component (either $\hat{q}$ or $\hat{p}$) that corresponds to the same basis in which he prepared the initial states, and makes corresponding homodyne detection.
Alice announces the positions and encoding information of the randomly selected check states. Bob decodes the corresponding states, calculates the quantum bit error rate (QBER), and if the QBER remains below the threshold, he proceeds to decode the secret information.

The displacement operator is defined as $\hat{D}(\alpha)=\exp(\alpha\hat{a}^{\dagger}-\alpha^{*}\hat{a})$, where $\alpha=\alpha_{\mathrm{R}}+i\alpha_{\mathrm{I}}$ is a complex displacement amplitude, with $\alpha_{\mathrm{R}}$ and $\alpha_{\mathrm{I}}$ denoting its real and imaginary parts, respectively. The displacement operator induces shifts of $\Delta q=\sqrt{2}\operatorname{Re}(\alpha)$ and
$\Delta p=\sqrt{2}\operatorname{Im}(\alpha)$ in phase space.
Accordingly, the displacement operation used in Step 4 for encoding corresponds to the $I$ or $Y$ operation, which simultaneously shifts the state by $(0$ or $\sqrt{\pi})$ along both the $\hat{q}$ and $\hat{p}$ quadratures, respectively. 
The error thresholds for the DBER and QBER used in the communication process are 5.6\% and 10.7\%, respectively.
In the simulation, the propagation of Gaussian errors is such that the variances of the $\hat{q}$ and $\hat{p}$ errors, initially $\sigma_{GKP}^2$ at preparation, become $c_{opt}\sigma_{GKP}^2$ after passing through quantum repeater, while logical X and Z errors $e_X$ and $e_Z$ will be introduced during the process.

\section{\label{sec.4}Security analysis and the final secrecy capacity}
In this section, we present a security analysis and derive the secrecy capacity of our QSDC protocol with GKP qubits. First, we consider the security of the error-correction operations performed at the repeater nodes. These GKP ancilla states enable error correction without revealing the encoded logical information. Furthermore, the CNOT gates are designed to extract the relevant error information while preserving the logical information encoded in the conjugate quadrature of the data mode.
Any attempt to extract information through measuring the ancilla state in a different basis would disturb the security check performed in the conjugate basis, thereby making the eavesdropping detectable.
More specifically, we know that, combining the Fourier-transform relation and the Poisson summation rule \cite{fourier}, the correspondence between the states in the Z basis and the X basis can be derived as follows.
\begin{align*}
	|0_L\rangle &= \sum_{n \in \mathbb{Z}} |q - 2n\sqrt{\pi}\rangle =\sum_{n \in \mathbb{Z}} \delta(q - 2n\sqrt{\pi}) |q\rangle\\
	&= \int \frac{dp}{\sqrt{2\pi}} \sum_{n \in \mathbb{Z}} e^{ip2n\sqrt{\pi}} |p\rangle \\
	&= \frac{1}{\sqrt{2}} \sum_{n \in \mathbb{Z}} \delta(p - n\sqrt{\pi}) |p\rangle
	= \frac{1}{\sqrt{2}} (|+_L\rangle + |-_L\rangle).
\end{align*}
For error correction in the $\hat{p}$ quadrature, the ancilla GKP state $\lvert 0_L \rangle$ is employed. Through the gate operation, the information contained in the $\hat{p}$ quadrature of $\lvert +_L \rangle$ or $\lvert -_L \rangle$ in data mode propagates to this ancillary mode. It can be observed that, in the $p$ quadrature, $\lvert 0_L \rangle$ is a superposition of even and odd multiples of $\sqrt{\pi}$.
Consequently, only the displacement for correction can be obtained by performing a modulo $\sqrt{\pi}$ operation on the syndrome measurement result. Simultaneously, the $q$ quadrature information (even multiples of $\sqrt{\pi}$) of the ancillary mode is transferred to the data mode, which does not corrupt the data mode.
Furthermore, if Eve attempts to eavesdrop using $\lvert +_L \rangle$ or $\lvert -_L \rangle$, although her attempt might succeed, it will inevitably disturb the information in the $\hat{q}$ of data mode, as it introduces an indeterminate displacement of odd or even multiples of $\sqrt{\pi}$, thereby causing a flip with a certain probability. The resulting bit error rate would exceed the predetermined threshold. 
From the above analysis, it can be seen that error correction in the quadratures only provides syndrome measurement results for feedback-based correction, which opens up the possibility for the design of our long-distance QSDC.

Second, in our protocol, continuous variables (CV) serve as the physical information carriers, while the communication transmits information as discrete logical bits. Consequently, we calculate the lower bound of the secrecy capacity according to Wyner’s wiretap channel theory \cite{Cs}.
\begin{equation}
    C_s = \max_{p_A} \{ I(A : B) - I(A : E) \}
\end{equation}
Here, $I(A:B)$ denotes the mutual information between Alice and Bob, $I(A:E)$ quantifies the maximum information that Eve can obtain from the communication, and $p_A$ is the probability that Alice applies her encoding displacement operation.

Our QSDC protocol requires two rounds of quantum transmission like Ref. \cite{DL04}. Specifically, we denote by BAB the round-trip process from Bob to Alice and then back to Bob, and by BA the one-way process from Bob to Alice. The mutual information between Alice and Bob, $I(A : B)$, is given by:
\begin{align}
	I(A : B) = Q^{BAB}[1 - h(E^{BAB})]
\end{align}
where the binary Shannon entropy $h(x)=-xlog(x)-(1-x)log(1-x)$; $E^{BAB}$ is the total quantum bit error rate (QBER) after a round-trip BAB, $Q^{BAB}$ is the overall signal gain of Bob, and we assume that
\(E^{BAB}=e_X^{BAB}=e_Z^{BAB}\),
with \(e_X^{BAB}\) and \(e_Z^{BAB}\) denoting the error rates in the X and Z bases, respectively.

    And $I(A : E)$ represents the information accessible to Eve about Alice's encoded message. In our protocol, the security analysis is restricted to collective attacks, following Ref. \cite{freeSpaceQSDC2020}. Specifically, Eve individually couples each transmitted quantum state to an ancillary state using the same unitary operation while storing her ancillary states in a quantum memory. After the encoding operation is performed and the encoded quantum state is transmitted back through the backward channel, Eve can perform an optimal joint measurement on her stored ancillary states and the returned encoded quantum states to extract information. 
    
    Based on this assumption, Qi et al. have derived the upper bound of information entropy that Eve can obtain \cite{practicalQSDC2019}. In our protocol, Bob prepares each of the four logical quantum states \(\{|0_L\rangle, |1_L\rangle, |+_L\rangle, |-_L\rangle\}\) with an equal probability of \(\frac{1}{4}\). The ideal GKP states prepared by Bob can be written as the density matrix:
	\begin{equation}
		\begin{aligned}
			\rho_1^B &= \frac{1}{4}(|0_L\rangle\langle 0_L| + |1_L\rangle\langle 1_L| + |+_L\rangle\langle +_L| + |-_L\rangle\langle -_L|) \\
			&= \frac{1}{2}(|0_L\rangle\langle 0_L| + |1_L\rangle\langle 1_L|).
		\end{aligned}
	\end{equation}
	Considering the practical preparation process, the prepared non-ideal GKP states undergo a bit-flip with probability \(p_{err}\) after measurement. Consequently, the prepared logical bit 0 can be effectively described as \((1-p_{err})\ket{0_L}\bra{0_L} + p_{err}\ket{1_L}\bra{1_L}\), and similarly for the logical bit 1. In this case, $\rho_1^B$ transforms into $\frac{1}{2}\{(1-p_{err}+p_{err})|0_L\rangle\langle 0_L| + (p_{err}+1-p_{err}) |1_L\rangle\langle 1_L|\}$, yet it remains identical to $\rho_1^B$.

	In the context of the collective attack, it is assumed that Eve applies the operation \(U_1\) to execute a joint operation on Bob's system and his own ancillary state \(|E\rangle\) \cite{practicalQSDC2019}, and
	\begin{equation}
		\rho_1^{BE} = U_1(\rho_1^B \otimes |E\rangle\langle E|)U_1^\dagger,
	\end{equation}
	where the effect of \(U_1\) can be represented as
	\begin{equation}
		\begin{aligned}
			U_1|0_L\rangle|E\rangle &= |0_L\rangle|E_{00}\rangle + |1_L\rangle|E_{01}\rangle = |\psi_1\rangle, \\
			U_1|1_L\rangle|E\rangle &= |0_L\rangle|E_{10}\rangle + |1_L\rangle|E_{11}\rangle = |\psi_2\rangle.
		\end{aligned}
	\end{equation}
	
	After Alice performs the encoding, we have
	\begin{equation}
		\begin{aligned}
			D(\alpha_0) U_1|0_L\rangle|E\rangle &= |0_L\rangle|E_{00}\rangle + |1_L\rangle|E_{01}\rangle = |\psi_1\rangle, \\
			D(\alpha_0) U_1|1_L\rangle|E\rangle &= |0_L\rangle|E_{10}\rangle + |1_L\rangle|E_{11}\rangle = |\psi_2\rangle, \\
			D(\alpha_1) U_1|0_L\rangle|E\rangle &= |1_L\rangle|E_{00}\rangle - |0_L\rangle|E_{01}\rangle = |\psi_3\rangle, \\
			D(\alpha_1) U_1|1_L\rangle|E\rangle &= |1_L\rangle|E_{10}\rangle - |0_L\rangle|E_{11}\rangle = |\psi_4\rangle.
		\end{aligned}
	\end{equation}
		
	Then the density matrix can be expressed as
	\begin{equation}
		\begin{aligned}
			\rho_1^{ABE} &= p_A D(\alpha_0) U_1(\rho_1^B \otimes |E\rangle\langle E|)U_1^\dagger 	D^\dagger(\alpha_0) \\
			&\quad + (1 - p_A) 	D(\alpha_1) U_1(\rho_1^B \otimes |E\rangle\langle E|)U_1^\dagger 	D^\dagger(\alpha_1) \\
			&= p_A \rho_1^{BE, 0} + (1 - p_A)\rho_1^{BE, 1} \\
			&= p_A \frac{1}{2}(|\psi_1\rangle\langle\psi_1| + |\psi_2\rangle\langle\psi_2|) \\
			&\quad + (1 - p_A)\frac{1}{2}(|\psi_3\rangle\langle\psi_3| + |\psi_4\rangle\langle\psi_4|),
		\end{aligned}
	\end{equation}
	where $\rho_1^{BE, 0}$ and $\rho_1^{BE, 1}$ denote the results of applying $D(\alpha_0)$ and $D(\alpha_1)$ to $\rho_1^{BE}$, respectively.
	
	According to the Holevo bound, the upper bound on \(I(A:E)\) can be expressed as \cite{holevo1973}
\begin{align}
	I(A:E) &\leq {} \max \Bigl\{ S(\rho_1^{ABE}) - p_A S(\rho_1^{BE, 0}) \notag \\ 
	& - (1 - p_A)S(\rho_1^{BE, 1}) \Bigr\},
\end{align}
	where \(S(\rho)\) is von Neumann entropy, and \(S(\rho) = -Tr(\rho \log_2 \rho) = -\sum_i \lambda_i \log_2 \lambda_i\). \(\lambda_i\) are the eigenvalues of \(\rho\).
	
	Given that the operations performed by Alice and Eve are unitary, the entropy satisfies \(S(\rho_1^{BE, 0}) = S(\rho_1^{BE, 1}) = S(\rho_1^B \otimes |E\rangle\langle E|) = 1\).  Since the non-zero eigenvalues of \(\rho^{ABE}\) are identical to those of its corresponding Gram matrix, we can determine its von Neumann entropy by solving the Gram matrix \cite{Neumannentropy}. 
	The $(i,j)$-th entry of this Gram matrix is given by $G_{i,j} = \sqrt{p_i p_j}\langle\psi_i|\psi_j\rangle.$
	Calculating the eigenvalues of the Gram matrix allows us to evaluate \(S(\rho_1^{ABE})\) \cite{solveeq}. When
	\begin{equation}
		\begin{aligned}
			&\langle E_{11}|E_{10}\rangle = \langle E_{00}|E_{01}\rangle = 0, \quad
			\langle E_{10}|E_{01}\rangle = -e_Z^{BA},  \\
			&\langle E_{00}|E_{11}\rangle = 1 - 2e_X^{BA} - 2e_Z^{BA},
		\end{aligned}
	\end{equation}
	\(S(\rho_1^{ABE})\) reaches the maximum value as
	\begin{equation}
		S(\rho_1^{ABE}) = 1 + h(e_Z^{BA} + e_X^{BA}).
	\end{equation}

Therefore, assuming that the bit error $e_Z^{BA}$ and phase error $e_X^{BA}$ of the GKP state are equal to $E^{BA}$, the maximum information accessible to Eve, denoted $I(A:E)$, is upper bounded as
\begin{align}
	I(A : E) \leq Q^{BAE} h(2E^{BA}),
\end{align}
where $E^{BA}$ is the DBER during the transmission from Bob to Alice, and $Q^{BAE}$ is the overall signal gain of Eve.

In our protocol, the measurement is performed by utilizing the continuous syndrome values. Therefore,
the practical homodyne detectors with efficiency $\eta_h$ will be considered in our protocol and incorporated into $\eta$ to ultimately affect the error correction performance. At the same time, the overall signal gain $Q^{BAB}, Q^{BAE}$ are both set to unity, while channel attenuation and detector efficiency affect the variance of the Gaussian distribution, ultimately influencing the probability of a logical error during GKP error correction.

Finally, the lower bound of secrecy capacity of our protocol is
$C_s = [1 - h(E^{BAB})] - h(2E^{BA})$.
Given $P^X_{err}$ and $P^Z_{err}$ for each repeater station, we can
compute the accumulated error rates for the Bob-Alice-Bob and Bob-Alice transmission channel, denoted $e_X^{BAB},e_X^{BA}$ and $e_Z^{BAB},e_Z^{BA}$ for X and Z errors, respectively.

\begin{figure}[htbp]
	\centering
	\includegraphics[scale=0.32]{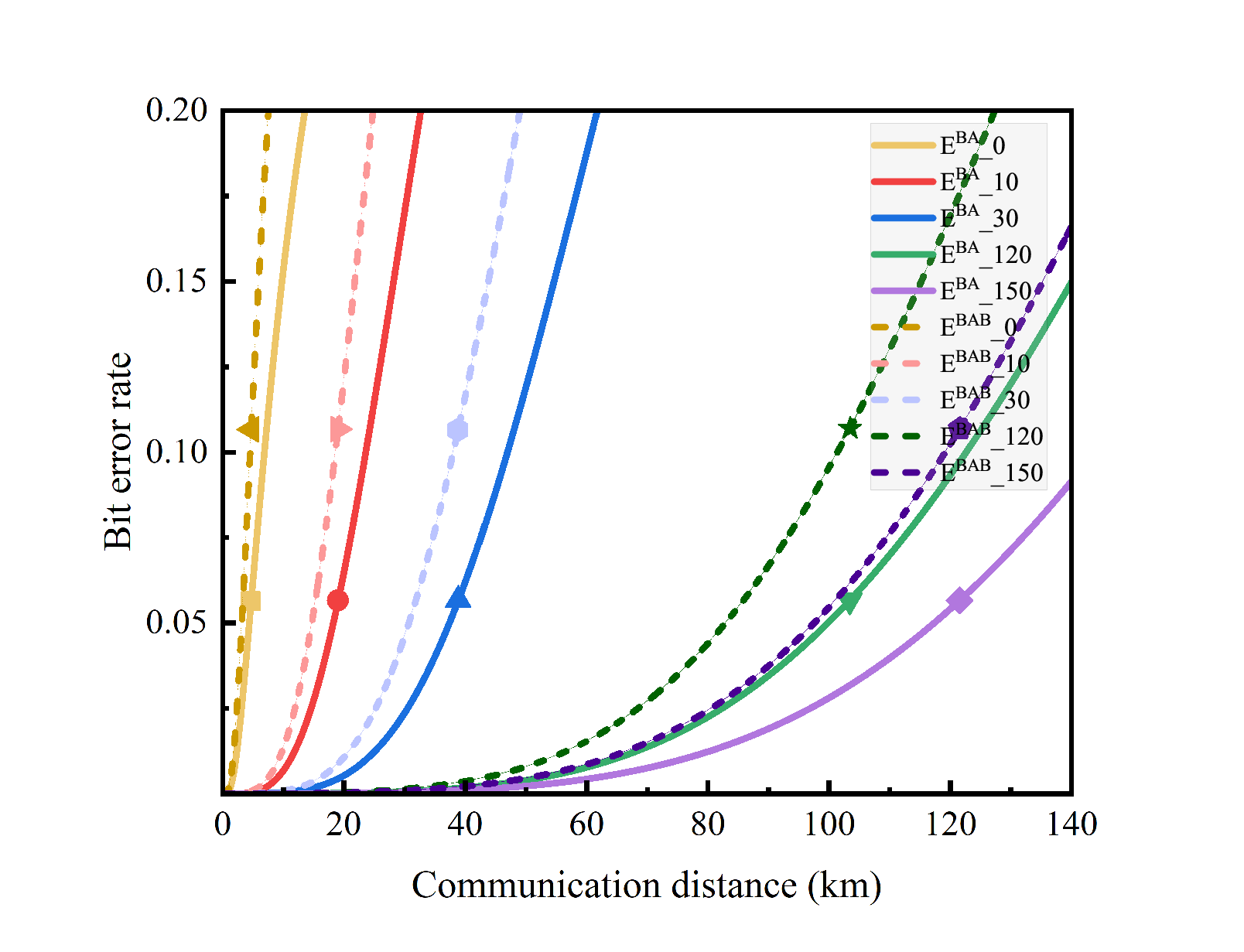}
	\caption{Bit error rate as a function of communication distance for the BA process (solid lines) and BAB process (dashed lines) with varying numbers of repeaters (0, 10, 30, 120, 150).  The markers on each curve denote the maximum tolerable bit error rate threshold at the maximum achievable communication distance.  Increasing the number of repeaters significantly extends the achievable communication distance under low bit error rates.}
	\label{FIG.999}
\end{figure}

In evaluating the accumulated logical error probabilities, we assume that logical errors occurring in different repeater nodes are statistically independent. Furthermore, the displacement noise in the $\hat{q}$ and $\hat{p}$ quadratures is modeled by two statistically independent Gaussian random variables, implying that the resulting logical X and Z errors are also independent.
Due to the symmetry of the square lattice, an even number of X or Z errors can be equivalent to no error. The probability $e_X^{BAB}, e_Z^{BAB}, e_X^{BA}$ and $ e_Z^{BA}$  can be approximated as follows.
\begin{align}
    &e_X^{BA} = e_Z^{BA}=\frac{1 - (1 - 2P^{X/Z}_{err})^{(NQR
    +1)}}{2}, \\
    &e_X^{BAB} = e_Z^{BAB}=\frac{1 - (1 - 2P^{X/Z}_{err})^{2(NQR
    +1)}}{2},
\end{align}
where NQR denotes the number of quantum repeaters, given by $(L_{all}/L)-1$.
The initial standard deviation of the GKP qubit is set to 0.05 , corresponding to a squeezing level of 23.01 dB. The fiber coupling efficiency and the efficiency of homodyne detector $\eta_0\eta_h$ is chosen to be 0.98, similar to the value $\eta_0\eta_h=99\%$ used in Ref. \cite{highdimensionGKP1}.

To characterize the DBER and QBER of our protocol, we plot their variations as functions of the total communication distance for different repeater configurations, as shown in Fig. \ref{FIG.999}. It can be observed that, for a given total communication distance, increasing the number of repeater nodes reduces both error rates. This improvement arises from the shorter elementary link lengths, which mitigate the accumulated displacement errors along the transmission channel. Furthermore, although the maximum achievable communication distance depends on the number of repeater nodes, the error-rate tolerance thresholds remain unchanged, as they are determined by the security requirements of the protocol. Specifically, the tolerable error-rate thresholds are approximately 5.6\% for the DBER in the first transmission stage, which serves for eavesdropping detection, and 10.7\% for the QBER after the two-stage transmission.

\section{\label{sec.5}simulation}

In our protocol, GKP qubits are integrated  to facilitate long-distance QSDC with built-in error correction. The GKP qubits can correct errors arising from channel loss and gate operations at each repeater node, while simultaneously resetting the variance. Thereby, long-distance QSDC is achieved through multiple error-correcting nodes.

\begin{figure}[htbp]
	\centering
	\includegraphics[scale=0.32]{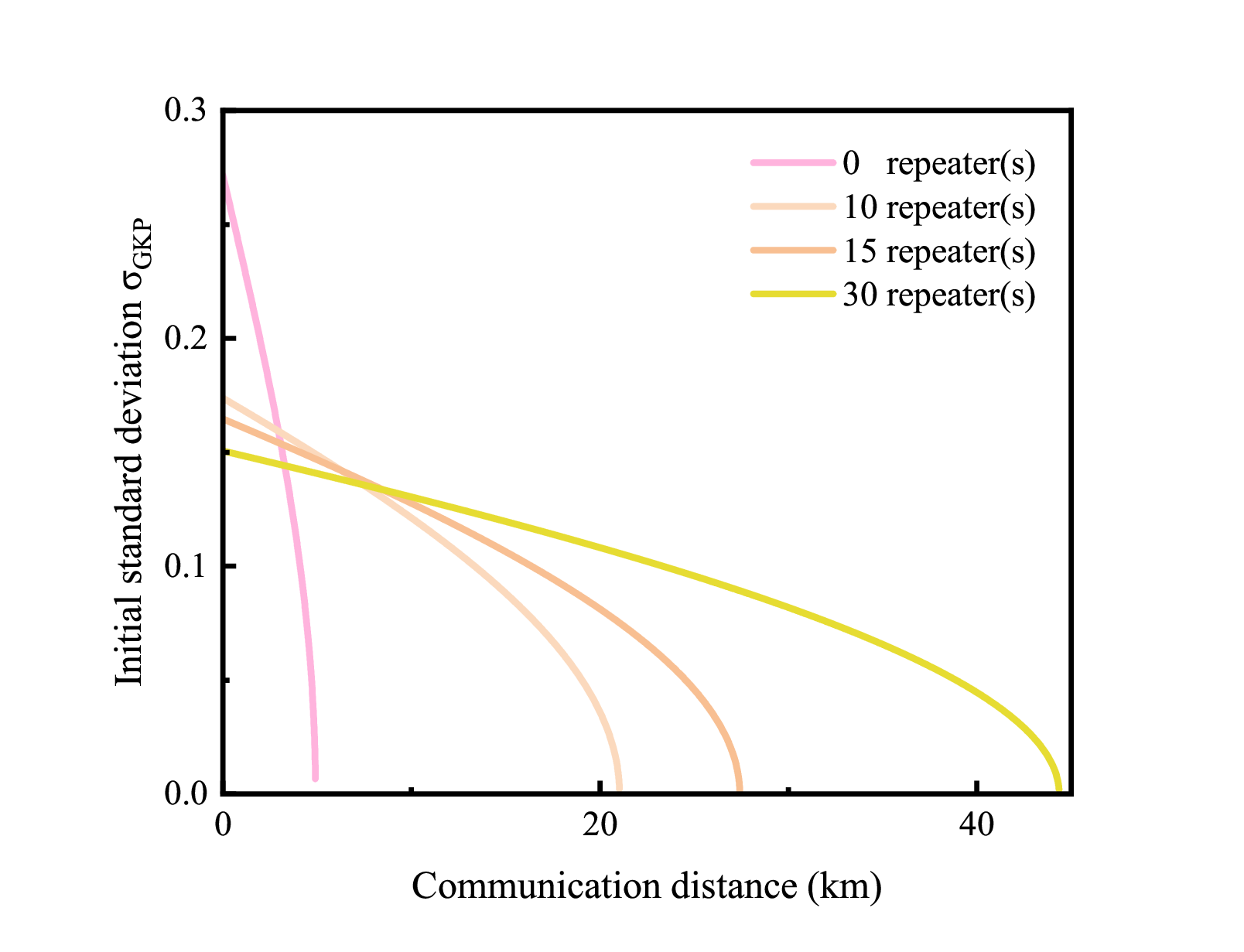}
	\caption{ The maximum tolerable initial standard deviation at different distances for $0, 10, 15,30$ repeaters with coupling and measurement efficiencies $\eta_0\eta_h$ = 0.98.}
	\label{FIG.888}
\end{figure}

First, the squeezing level of the initially prepared finite-squeezing GKP states is a key parameter governing the error-correction performance of our protocol. It is related to the initial standard deviation $\sigma_{\mathrm{GKP}}$ through $s=-10\log_{10}(2\sigma_{\mathrm{GKP}}^2)$. To investigate the influence of the initial-state quality on the achievable communication distance, Fig. \ref{FIG.888} illustrates the maximum tolerable initial standard deviation as a function of the communication distance for different repeater configurations, with 0, 10, 15, and 30 repeater nodes. It can be seen that the introduction of repeaters can effectively improve the system’s tolerance to the initial standard deviation at the same distance.
Based on this comparison, we choose $\sigma_{\mathrm{GKP}}=0.05$, corresponding to a squeezing level of 23.01 dB, as the initial standard deviation for the subsequent calculations. This choice provides a common finite-squeezing benchmark for comparing different repeater configurations and allows us to evaluate the improvement in communication distance while maintaining the same initial-state quality.

\begin{figure}[htbp]
	\centering
	\includegraphics[scale=0.33]{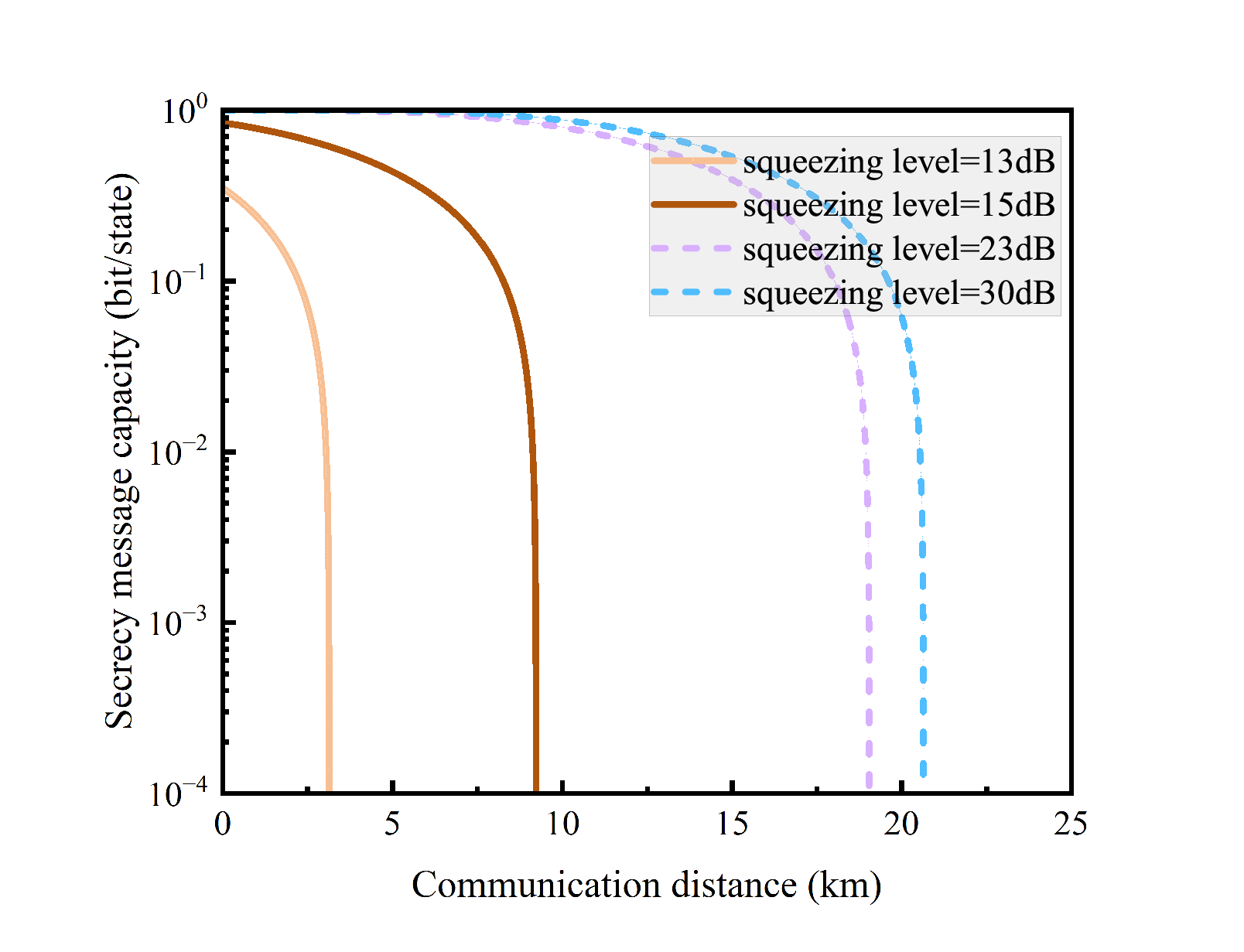}
	\caption{The secrecy message capacity as a function of communication distance under various squeezing levels, with the number of repeaters fixed at 10, coupling and measurement efficiencies $\eta_0\eta_h$ = 0.98.}
	\label{FIG.99}
\end{figure}

To further demonstrate,
Fig. \ref{FIG.99} illustrates the relationship between the secrecy message capacity and the communication distance for different squeezing levels, with the number of repeaters fixed at $10$.
As the squeezing level increases, error correction performance improves, leading to a longer communication distance. However, with further increases in the squeezing level, the performance improvement gradually diminishes, and the additional benefit in mitigating channel loss becomes marginal. 
Consequently, our protocol balances performance with practical constraints and adopts a squeezing level of 23 dB.

\begin{figure}[htbp]
\centering
\includegraphics[scale=0.37]{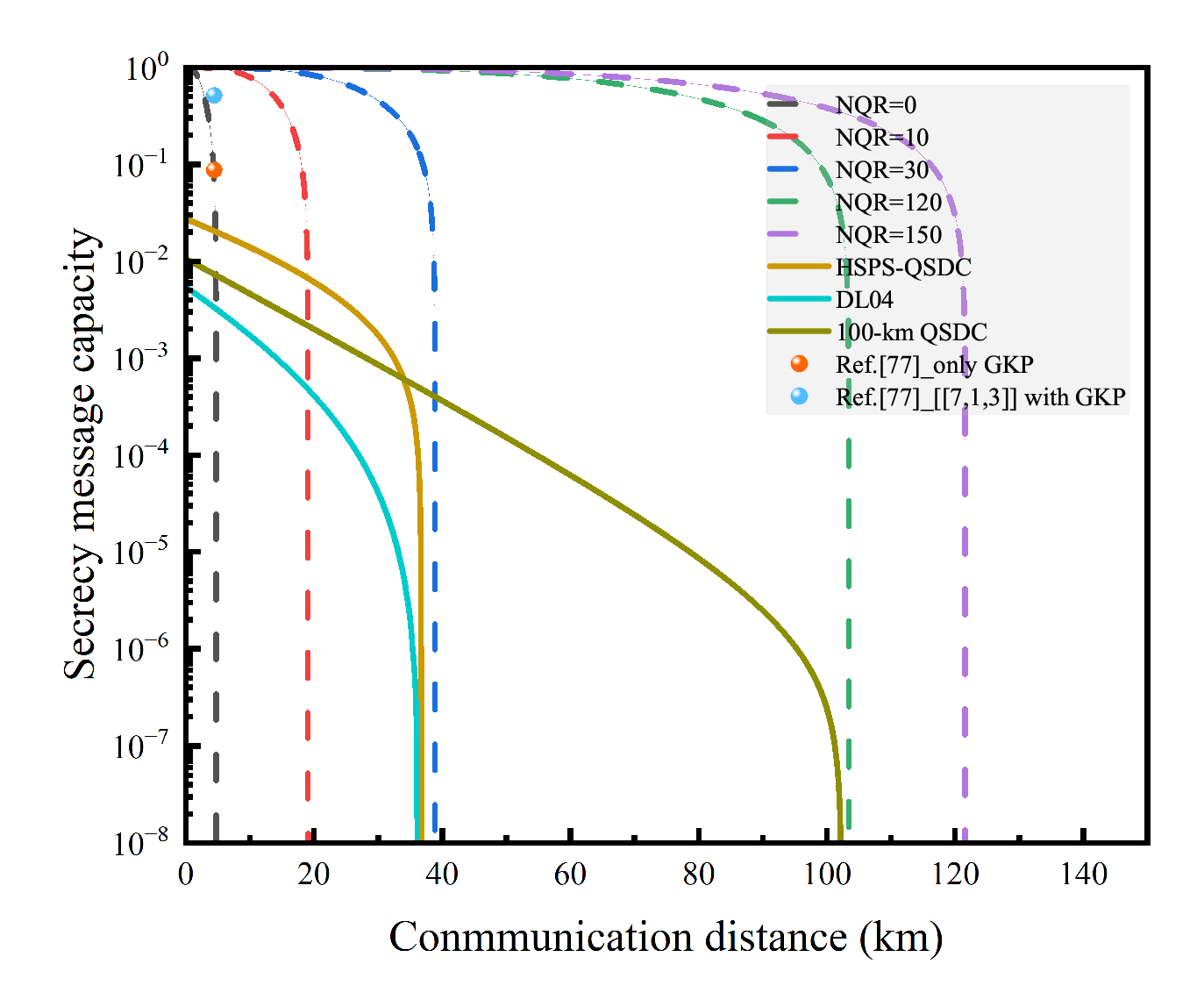}
\caption{The secrecy message capacity as a function of communication distance for the proposed protocol with different numbers of quantum repeaters $NQR =0, 10, 30, 120, 150$, coupling and measurement efficiencies $\eta_0\eta_h$ = 0.98, and squeezing level of 23 dB,  along with those of the HSPS-QSDC \cite{passiveDecoyQSDC2024}, DL04 \cite{freeSpaceQSDC2020}, 100-km QSDC protocol \cite{experiment3}, and the reference points from Ref. \cite{GKPQldpc} (GKP-only and GKP–QLDPC concatenated scheme with a  $\llbracket 7,1,3 \rrbracket$ outer code).}
\label{FIG.5}
\end{figure}

Our protocol adopts the third-generation quantum repeater architecture, which cannot rely on entanglement purification. Instead, it relies entirely on the intrinsic error‑correcting capability of the encoded qubits.
For comparison, we consider several well-established QSDC protocols, including the HSPS-QSDC protocol \cite{passiveDecoyQSDC2024}, the DL04 protocol \cite{freeSpaceQSDC2020} and a 100-km QSDC protocol based on time-bin and phase quantum states with ($e_x=0.001$) and ($e_z=0.0003$) \cite{experiment3}. Both the HSPS-QSDC and DL04 protocols employ decoy-state techniques. In addition, two reference points from Ref. \cite{GKPQldpc}, corresponding to the GKP-only and the GKP–QLDPC concatenated scheme with a $\llbracket 7,1,3 \rrbracket$ outer code are included for comparison. The secrecy message capacity as a function of communication distance for these protocols is presented in Fig. \ref{FIG.5}.

\begin{table}[t]
	\centering
	\caption{Maximum communication distances for different numbers of quantum repeater nodes.}
	\label{tab:max_distance}
	\small
	\renewcommand{\arraystretch}{1.25}
	\setlength{\tabcolsep}{12pt}
	\begin{tabular}{cc}
		\toprule
		\shortstack{Number of\\repeaters (NQR)}
		&
		\shortstack{Maximum distance\\$L_{\max}$ (km)} \\
		\midrule
		0   & 4.70 \\
		10   & 19.04 \\
		30  & 38.88 \\
		120  & 103.45 \\
		150  & 121.59 \\
		2750 & 1010.37 \\
		\bottomrule
	\end{tabular}
	\label{talbeL}
\end{table}

Without quantum repeaters (\(NQR=0\)), the proposed protocol achieves a maximum communication distance of only 4.70 km, primarily limited by fiber attenuation and displacement noise. Introducing quantum repeaters substantially enhances the long-distance transmission performance. By dividing the transmission channel into shorter elementary links and performing error correction at the repeater nodes, the proposed architecture mitigates the effects of channel loss and displacement errors, thereby extending the achievable communication distance. As shown in Fig. \ref{FIG.5}, the maximum communication distance increases progressively with the number of repeater nodes. Specifically, it reaches 19.04 km with 10 repeater nodes and 38.88 km with 30 repeater nodes, exceeding the transmission ranges of both the HSPS-QSDC \cite{passiveDecoyQSDC2024} and DL04 protocols \cite{freeSpaceQSDC2020}. With 120 repeater nodes, the achievable distance further extends to 103.45 km, surpassing the 100-km QSDC benchmark \cite{experiment3}. Increasing the number of repeater nodes to 150 extends the maximum distance to 121.59 km. More remarkably, with 2750 repeater nodes, the proposed protocol achieves a communication distance of 1010.37 km, exceeding 1000 km and representing an approximately 215-fold increase over the repeater-free configuration. The maximum communication distances for different repeater configurations are summarized in Tab. \ref{talbeL}.

For the same pure-loss channel model considered here, the only-GKP scheme \cite{GKPQldpc} performs similarly to our repeaterless case. 
The concatenated code \cite{GKPQldpc}  achieves lower logical error rates and higher secrecy capacity than the GKP-only scenario in a certain regime. 
However, this advantage is subject to the error threshold of the outer code. In the point-to-point direct transmission scenario considered in that protocol, the effective displacement-noise variance increases with distance, potentially driving the underlying GKP error rate beyond the correctable threshold. Consequently, the performance advantage of concatenated coding gradually diminishes and may eventually disappear.
Overall, these results demonstrate the potential of the proposed repeater-assisted QSDC architecture for ultra-long-distance quantum secure communication.

Based on the above analysis, for a fixed transmission distance, increasing the number of repeater nodes enhances the ability to mitigate channel loss, thereby enabling high-rate communication. To more effectively evaluate the performance of the QSDC protocol, we introduce a cost function.
Like Ref. \cite{concatenated1,highdimensionGKP1}, each repeater performs GKP error correction on the data mode in both conjugate quadratures. With two ancilla storage modes operated in alternation, the data mode is stored for two time steps and each ancilla for one, giving a resource cost
$t_{GKP}=4$ storage-mode time steps.
Additionally, we consider the process in which Bob initially sends and finally receives the GKP qubits to be incorporated into these time steps required for a complete repeater structure due to the periodic condition.

Similar to Refs. \cite{concatenated1,highdimensionGKP1},  we  define the cost function as:
\begin{align}
	C(L_{all}, NQR)&=\frac{\text{Resources used per km}}{\text{secrecy message capacity}}\notag \\
	&=\frac{(2NQR+3) t_{GKP}}{C_s L_{all}}.
\end{align}
Since this protocol requires two rounds of transmission using the framework in Ref. \cite{DL04}, the total number of time steps is $(2NQR+3)t_{GKP}$. In the simulation, we assume that at most one repeater node can be configured per $0.25$ km (i.e., $250$ m). For a given total transmission distance $L_{\text{all}}$, the maximum allowable number of repeater nodes is therefore $\min\!\left(N, \left\lfloor L_{\text{all}} / 0.25 \right\rfloor\right)$, where $N$ denotes a given upper bound. We aim to identify the optimal number of repeater nodes $NQR{opt}$ that minimize the cost function under different transmission distances.

\begin{figure*}[htbp]
	\centering
	\includegraphics[scale=0.3]{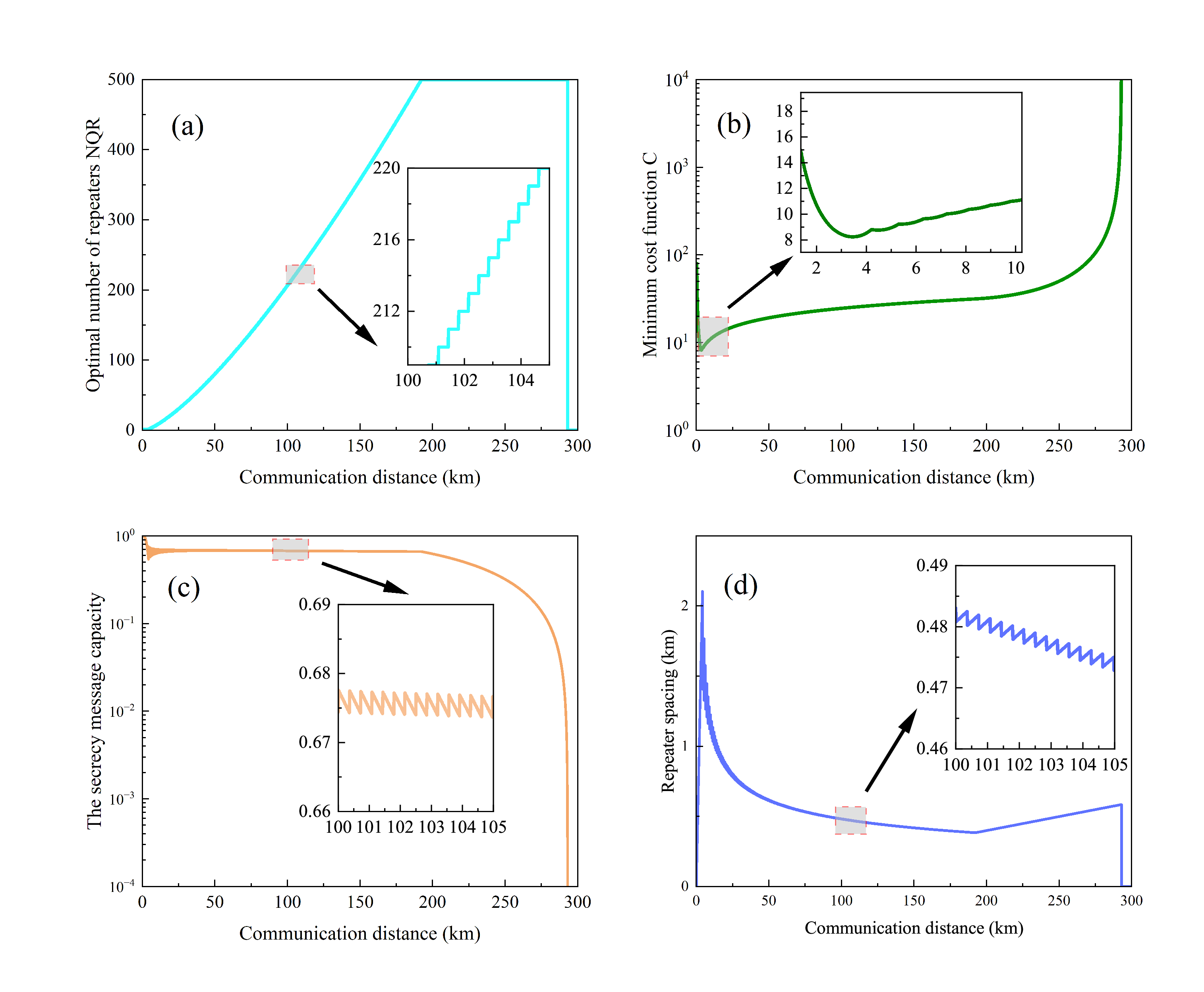}
	\caption{Performance analysis as a function of total communication distance for $N$ = 500 with coupling and measurement efficiencies $\eta_0\eta_h$ = 0.98, and squeezing level of 23 dB. (a) the optimal number of repeater nodes, (b) the minimum cost function, (c) the secrecy message capacity with optimal repeater configuration, and (d) corresponding repeater spacing under different transmission distance.}
	\label{FIG.6}
\end{figure*}

Fig. \ref{FIG.6} illustrates the optimal number of repeater nodes, the minimum cost function, the corresponding secrecy
message capacity and repeater spacing under different transmission distances for $N = 500$. The relevant parameters are listed in Tab. \ref{tab:params}. By minimizing the cost function, we can identify the optimal repeater configuration that balances the secrecy capacity and resource utilization for communication distances. As long as the configured number of repeaters does not exceed the maximum allowable value, as the communication distance increases, more repeaters are required to overcome channel loss, as depicted in Fig. \ref{FIG.6}a.

\begin{table}[htbp]
	\centering
	\caption{The parameters adopted in the simulations. $\eta_0,\eta_h$ are the fiber coupling efficiency and the efficiency of homodyne detector, $L_{\text{att}}$ is the fiber attenuation length (in km), $\sigma_{\text{GKP}}$ is the standard deviation of the GKP state, $\alpha$ is the channel's loss rate, and $t_{\text{GKP}}$ is the time-step cost per single GKP repeater.}
	\label{tab:params}
	\begin{tabular*}{\columnwidth}{c@{\extracolsep{\fill}}ccccc}
		\toprule
		Parameter & $\eta_0\eta_h$ & $L_{\text{att}}$ & $\sigma_{\text{GKP}}$ & $\alpha$ & $t_{\text{GKP}}$ \\
		\midrule
		Value & 0.98 & 22 & 0.05 & 0.2 & 4 \\
		\bottomrule
	\end{tabular*}
\end{table}

Fig. \ref{FIG.6}b illustrates that, since the number of repeaters must be an integer, when the optimal configuration $NQR_{opt}$ is exactly achieved at a specific communication distance, the cost function will change as the distance increases, owing to the trade-off between resource consumption per kilometer and the achievable secrecy capacity, without yet transitioning to the configuration $NQR_{opt}+1$. This indicates the presence of a local minimum, which corresponds to the optimal communication distance for a fixed $NQR_{opt}$.
During this process, the secrecy message capacity continuously decreases (Fig. \ref{FIG.6}c), while the optimal repeater spacing increases (Fig. \ref{FIG.6}d). Once the configuration transitions to $NQR_{opt}+1$, the cost function changes again.

For communication distances over which the optimal repeater number remains below its maximum allowed value, the secrecy message capacity remains nearly constant at approximately 0.67 bit/state (Fig. \ref{FIG.6}c), whereas the optimal repeater spacing exhibits an overall decreasing trend, indicating progressively denser repeater deployment (Fig. \ref{FIG.6}d).
As shown in Fig. \ref{FIG.6}b, the overall trend of the cost function is an initial decrease followed by a continuous increase within the communication distance range of 0 km to 300 km, exhibiting a global minimum within the considered range.
This minimum occurs at $L_{\mathrm{all}}=3.46$ km, corresponding to one repeater node and a secrecy message capacity of 0.70 bit/state.

Although this minimum-cost point corresponds to a relatively short communication distance, the cost function increases only gradually over a broad range of longer distances, indicating that extending the communication distance does not necessarily incur a substantial increase in cost. Thus, for our protocol, the minimum-cost configuration is not necessarily the most suitable operating point for long-distance communication, and operating away from the global minimum can provide a favorable trade-off between communication distance and resource cost. 
However, when the required number of repeaters exceeds the maximum allowable limit, the secrecy message capacity decays exponentially as the communication distance increases, and the cost function rises continuously.

Beyond identifying the minimum-cost operating point, this analysis provides a quantitative criterion for optimizing repeater deployment at different communication distances. Such a criterion is particularly useful for future extensions to cascaded repeater architectures, where the cost and transmission performance of multiple concatenated segments must be jointly optimized.

\section{\label{sec.6}discussion}
The challenge of our QSDC protocol ultimately lies in its experimental realization.
Firstly, it is feasible to implement the gate operations and homodyne detection \cite{cv-operation1,cv-operation2,cv-operation3}, which are employed in the error-correction process. At the same time, like Ref. \cite{concatenated1}, the CV quantum memories are necessary in our protocol, and superconducting microwave cavities provide an attractive platform for the preparation, storage, and error correction of GKP states \cite{superconduct2}. However, their use in fiber-based communication requires transduction \cite{transduction1,transduction2} between telecom optical modes and microwave modes with high efficiency and low added noise. This optical-microwave hybrid architecture as a promising approach and has already been accounted for in our cost-function evaluation.
Another possible route is an all-optical online implementation, in which error correction is performed directly on flying optical GKP modes with fiber delays \cite{optical_experiment1,qsdcreview,fiberLoopMemory2025} and coupled optical ancillary GKP states. This approach avoids the need for long-lived matter-based quantum memories. Its main challenges are the generation of high-quality optical GKP states \cite{opticalGKP,2026opt}, fast feed-forward control, and precise synchronization between data and ancillary modes. At present, the experimentally demonstrated squeezing level of optical GKP-like states remain below the parameter regime required in our long-distance simulations. Therefore, improving the quality of optical GKP-state generation remains a key requirement for this implementation.

In addition, finite SUM-gate fidelity, imperfect feed-forward displacement, memory decoherence, and mode mismatch may all increase the effective displacement-noise variance. Importantly, this two-round QSDC protocol needs to provide additional temporary storage during the security-checking and message-encoding procedure at Alice's station. 
These realistic imperfections and architectural requirements should be incorporated in future analyses.

\section{\label{sec.7}conclusion}

In this work, we have proposed a GKP‑based long-distance QSDC protocol, where the secret message can be transmitted directly, leveraging the intrinsic error-correcting capabilities of the GKP code. Moreover, benefiting from detecting eavesdropping in real time during communication, this protocol  provides an effective way to protect the lives of communication personnel, avoiding decapitation strikes in modern warfare caused by exposing communication positions. Combined with the third-generation quantum repeater,  the protocol effectively overcomes photon loss and accumulated noise. Simulation results demonstrate that the proposed protocol can achieve secure communication over 1000 km with a squeezing level of 23 dB.
By balancing resource consumption and the secrecy message capacity, we have identified optimal repeater configurations for different transmission distances.  We provide optimization under the upper limit of 500 repeater nodes over 300 km. 
The resulting optimization criterion provides guidance for repeater deployment at different communication distances and offers a basis for future extensions to other repeater architectures.
Our work highlights the potential of GKP qubits as a practical building block for future long-distance QSDC and quantum networks, particularly in scenarios requiring both security and robustness against channel loss. Future research may explore the integration of this protocol with advanced quantum repeater architectures, optimization for finite-size effects, and experimental realization using existing photonic platforms.

\begin{acknowledgments}
This work is supported by the National Natural Science Foundation of China under Grants  Nos. 12574393, 92365110 and 12634015.
\end{acknowledgments}


\end{document}